\documentclass[12pt]{article}

\newcommand{\ket}[1]{\left|#1\right\rangle}
\newcommand{\pe}[1]{\int d^3x\left(#1\right)}

\newcommand{\emn}{\varepsilon^{\mu\nu\lambda}}
\newcommand{\con}[2]{\left[#1,#2\right]}

\newcommand{\eref}[1]{(\ref{#1})}
\newcommand{\vphi}{\varphi}
\newcommand{\db}[2]{\left\{#1,#2\right\}^*}
\renewcommand{\sup}[1]{^{(#1)}}
\newcommand{\dxy}{\delta\sup 2(\vec x-\vec y)}
\newcommand{\dtxy}{\delta\sup 3(\vec x-\vec y)}
\begin{document}
\begin{flushright}hep-th/0206082\\UCVFC-DF-16-2001
\end{flushright}
\vskip 1truein
\begin{center}
{\Huge{A Geometric Approach to Massive $p$ form Duality}}\\
{\bf{P\'{\i}o J. Arias{${}^{a, b, }$}\footnote{parias@fisica.ciens.ucv.ve},
Lorenzo Leal{${}^{a,c, }$}\footnote{lleal@fisica.ciens.ucv.ve} and
J.C. P\'erez-Mosquera{${}^{a,
}$}\footnote{jcperez@fisica.ciens.ucv.ve}}}\\
${}^a${\it Grupo de Campos y Part\'{\i}culas, Departamento de F\'{\i}sica, Facultad de Ciencias, 
Universidad
Central de Venezuela, AP 47270, Caracas 1041-A, Venezuela} \\
${}^b${\it Centro de Astrof\'{\i}sica Te\'orica, Facultad de Ciencias,
Universidad de Los Andes, La Hechicera, M\'erida 5101,
Venezuela}\\ ${}^c$ {\it Departamento de F\'{\i}sica Te\'orica,
Universidad Aut\'onoma de Madrid, Cantoblanco 28049, Madrid,
Spain}
\end{center}


\begin{abstract}
Massive theories of abelian $p$-forms are quantized in a generalized
path-representation that leads to a description of  the phase space
in terms of a pair of dual non-local operators analogous to the Wilson
Loop and the 't Hooft disorder operators. Special atention is devoted to
the study of the duality between the Topologically Massive and the Self-Dual
models in $2+1$ dimensions. It is shown that these models share a geometric
representation in which just one non local operator suffices to describe
the observables. 
\end{abstract}

\newpage

\section{Introduction}
As it is known, the electric-magnetic duality of free Maxwell
theory may be seen as a particular case of a duality
tansformation relating different abelian gauge theories of
arbitrary rank in the appropriate space time
dimensions\cite{Townsend:1981nu,DFNS}. In $D$ dimensions, the
ranks $p_1$, $p_2$ of the generalized potentials describing the
dual abelian theories must obey the equality $p_1 + p_2 = D-2$.
For example, in four dimensions, Maxwell theory is self-dual,
while the second rank gauge theory is dual to the massles scalar
field.

In reference\cite{LL} it was shown that the ``electric-magnetic''
duality of abelian gauge theories allows to describe their
physical phase space in terms of a pair of non-local observables
that are dual in the Kramers-Wannier sense\cite{Kramers:1941kn}.
The algebra that they obey results to be invariant under spatial
diffeomorphisms. This topological algebra, the Dual Algebra (DA),
admits a realization in terms of operators acting on functionals
that depend on extended objects, inasmuch as the dual operators
themselves. For instance, in the case of Maxwell theory in four
space time dimensions, the dual operators are the Wilson Loop and
the 't Hooft disorder operator\cite{tHooft}. Both operators
depend on closed spatial loops, and may be realized on a
loop-dependent Hilbert space (see section \ref{sec2}). The DA of
the three and four dimensional Maxwell theory had been previously
analized\cite{AlgNL-Maxwell}, due to their close relation with the
Yang-Mills field. Furthermore, non local operators that obey
commutation relations of the DA type have been used to quantize
topological excitations in interacting field
theories\cite{Marino:1992rq}.

In this paper we discuss how the ideas of reference\cite{LL} can
be extended to the conventional (i.e., non topological) Abelian
massive theories in arbitrary dimension, and to the Self- Dual
\cite{TPN} and Topologically Massive theories\cite{DJT} in 2+1
dimensions, which are known to be dual to each other\cite{DJ}. In
all the cases, the program that we develop is as follows: one
starts from a first order master lagrangian that encodes the dual
theories simultaneously. We take this master lagrangian to be of
the St\"ueckelberg form~\cite{St}, in order to maintain gauge
invariance even in the massive case. The master theory is then
quantized within the Dirac scheme\cite{Dirac}, and the phase
space is taken into account by choosing non local operators that
encode all the gauge invariant content of the original canonical
operators. The algebra obeyed by these dual operators is then
studied and realized onto an appropriate set of functionals.

We shall see that the DA of massive Abelian theories is also
characterized by a topological quantity, namely, the intersection
number between the extended objects that support the non-local
dual operators. This contrasts with the masless case, where the
DA is governed by the linking number of the closed extended
objects that enter in the construction of the dual
operators\cite{LL,AlgNL-Maxwell}. This and other differences
between both cases are studied.

The case of the Self-Dual  and Topologically Massive theories
presents several interesting peculiarities, regarding the DA
study. Perhaps the more relevant one is that instead of a pair of
Wilson Loop  operators, as in both the massless and the
conventional massive theories, it suffices with only one
non-local operator to describe the gauge invariant content of the
theory. Consecuently, this operator has to play both the
``coordinate'' and ``momentum'' roles. As we shall see, this
feature has an interesting geometrical counterpart when the
non-local operator is realized in a path-dependent Hilbert space. In other
direction the Proca model in 2+1 dimensions is equivalent to two non-interacting
Self-Dual models with opposite spins. 

The paper is organized as follows. In section \ref{sec2} we
review the massless case, following reference\cite{LL}, focusing
mainly in the study of Maxwell theory in four dimensions. In
section \ref{sec3} the DA of the Proca model in three dimensions
is considered. Section \ref{sec4} is dedicated to the study of the
Self-Dual and Topologically Massive theories. Some concluding
remarks are given in the last section. In the Appendix we
summarize the generalization of the study presented in section
\ref{sec3}, to the case of forms of arbitrary rank in arbitrary
dimension.
\section{Maxwell theory}\label{sec2}
  Let us summarize the results of reference\cite{LL} regarding Maxwell
theory. The starting point is the first order Lagrangian density
\begin{equation}
{\cal L}=\frac 12\epsilon^{\mu\nu\lambda\rho}\partial_\mu A_\nu
B_{\lambda\rho}-\frac 14\left(B_{\lambda\rho}+\partial_\lambda
C_\rho-\partial_\rho C_\lambda\right)\left(B^{\lambda\rho}+\partial^\lambda
C^\rho-\partial^\rho C^\lambda\right),
\end{equation}
which is invariant under the simultaneous gauge transformations
\begin{eqnarray}
\delta A_\mu&=&\partial_\mu\Lambda,\\
\delta B_{\lambda\rho}&=&\partial_\rho\xi_\lambda-\partial_\lambda\xi_\rho,
\\
\delta C_\rho&=&\xi_\rho + \partial_\rho\xi.\label{deltaC}
\end{eqnarray}
Equation \eref{deltaC} shows that the field $C_\rho$ is pure
gauge. Its presence just serves to enforce gauge invariance. When
this field is gauged away in (1), the equations of motion become
\begin{eqnarray}
\epsilon^{\mu\nu\lambda\rho}\partial_\mu A_\nu&=&B^{\lambda\rho},\label{EdMA}\\
\epsilon^{\mu\nu\lambda\rho}\partial_\mu B_{\lambda\rho}&=&0.\label{EdMB}
\end{eqnarray}
Substituting $B_{\lambda\rho}$ from \eref{EdMA} into the master lagrangian
(with $C_\rho=0$), one finds the standard Maxwell lagrangian. If, instead,
one solves equation \eref{EdMB} locally
\begin{equation}
B_{\lambda\rho}=\partial_\lambda\tilde A_\rho-\partial_\rho\tilde
A_\lambda,
\end{equation}
and substitutes the above expression into the equation (1) (again with $C_\rho=0$)
we obtain (after an integration by parts) the ``dual'' lagrangian density
\begin{equation}
\tilde L=-\frac 14 F_{\mu\nu}(\tilde A) F^{\mu\nu}(\tilde A),
\end{equation}
in correspondence with the fact that in $D=4$ Maxwell theory is self-dual.

The canonical analysis may be summarized as follows. There are
three secondary first class constraints
\begin{eqnarray}
\psi&=&-\frac 12 \epsilon^{ijk}\partial_i B_{jk}\approx0,\label{ligadura1}\\
\theta^i&=&-(\pi_C^i-\epsilon^{ijk}\partial_j A_k)\approx0\label{ligadura2}\\
\theta&=&-\partial_i\pi_C^i\approx0\label{ligadura3}
\end{eqnarray}
where $\pi_c^i$ is the momentum canonically conjugate to $C_i$.
We are taking $g_{\mu\nu} = \hbox{diag}(1,-1,-1,-1)$. These constraints
are reducible ($\partial_i\theta^i-\theta=0$) and appear
associate, respectively, to $A_0$, $B_{0i}$ and $C_0$ as their
Lagrange multipliers.

The fields $A_i$ and $\frac 12\epsilon^{ijk}B_{jk}$ are mutually conjugate
\begin{equation}
\left[A_k(\vec x),\frac 12\epsilon^{lij}B_{ij}(\vec y)\right]=i\delta_k^l\dtxy,
\end{equation}
as can be seen from the first order $BF$ term in the master lagrangian
(see \cite{Faddeev-Jackiw}). $\psi$, $\theta^i$ and $\theta$ generate
the gauge transformations for $A_i$, $B_{ij}$, $C_i$ and $\pi_C^i$.
The gauge transformations for the remaining fields are obtained imposing
the gauge invariance on the extended action, taking into account the
reducibility of the first class constraints.

On the physical sector, the Hamiltonian reduces to
\begin{equation}
H=\int d^3\vec x\frac 12\left({\cal B}^i{\cal B}^i+{\cal E}^i{\cal E}^i\right),
\end{equation}
with the magnetic and electric fields given, respectivelly, by
\begin{eqnarray}
{\cal B}^k&\equiv&\epsilon^{ijk}\partial_j A_i,\\
{\cal E}^i&\equiv&\frac 12\epsilon^{ijk}\left(B_{jk}+F_{jk}(C)\right).
\end{eqnarray}

The gauge invariant combinations of the operators appearing in the above
expressions indicate which are the non-local dual operators we are interested
in. They are the Wilson loop
\begin{equation}
W(\gamma)=\exp\left(i\oint_\gamma dy^iA_i(\vec y)\right),\label{WL-Maxwell}
\end{equation}
with $\gamma$ a closed spatial path, and the operator
\begin{equation}
\Omega(\Sigma,\Gamma)=\exp{\left(i\oint_\Gamma dy^iC_i(\vec y)\right)}
\exp{\left(i\int_\Sigma d\Sigma_k\epsilon^{kij}B_{ij}\right)},
\label{WD-Maxwell}.
\end{equation}
wich depends on the spatial open surface $\Sigma$ whose boundary is
$\Gamma$. In virtue of the constraint \eref{ligadura1}, one has
\begin{equation}
\Omega\left(\Sigma_{\hbox{closed}}\right)\ket{\psi_{\hbox{physical}}}=
\ket{\psi_{\hbox{physical}}},
\end{equation}
i.e. $\Omega$ does not depend on the surface $\Sigma$, but only on its
boundary $\Gamma$. The algebra obeyed by the dual operators (the DA) is
given by
\begin{equation}
W(\gamma)\Omega(\Gamma)=e^{i{\cal L}(\gamma,\Gamma)}
\Omega(\Gamma)W(\gamma).\label{AlgDual}
\end{equation}
where the quantity
\begin{equation}
{\cal L}(\gamma,\Gamma)=\frac 1{4\pi}\oint_\gamma dx^i\oint_\Gamma
dy^j\epsilon_{ijk}\frac{(\vec x-\vec y)^k}{|\vec x-\vec y|^3},
\end{equation}
measures the Gauss linking number between $\gamma$ and $\Gamma$,
which are closed curves in $R^3$, and is a topological object, since
it does not depend on the metric properties of the space.

The operator $\Omega (\Gamma)$ results to be the ``dual'' Wilson loop, i.e.
the contour integral of the dual potential $\tilde A$ along
$\Gamma$\cite{AlgNL-Maxwell}. It must be noticed, however, that these results
are obtained from a formulation that does not include this potential as a
lagrangian variable, which would be redundant.

The DA \eref{AlgDual} is fulfilled if the operators are defined to act onto
loop dependent functionals $\Psi(\gamma)$ as
\begin{eqnarray}
W(\gamma)\Psi(\gamma_1)&=&\Psi({\gamma}\circ{\gamma_1}), \label{RealW} \\
\Omega(\Gamma)\Psi(\gamma_1)&=&e^{-i{\cal L}(\Gamma,\gamma_1)}\Psi(\gamma_1).
\label{RealO}
\end{eqnarray}
Here ${\gamma}\circ{\gamma}'$ denotes the Abelian group of loops
product\cite{GT-em,NC:Max}. It is worth recalling that an Abelian
loop is an equivalence class of closed curves, defined as
follows. The curves $\gamma_1$ and $\gamma_2$ are equivalent if
their form factors $T^i(\vec x,\gamma_1)$ and $T^i(\vec x,\gamma_2)$, with
\begin{equation}
T^i(\vec x,\gamma)\equiv\int_\gamma dy^i\dtxy,
\end{equation}
are equal. With this definition it is easy to see that the usual composition
of curves is lifted to a group product.

The electric and magnetic fields may be obtained from $W(\gamma)$ and
$\Omega(\Gamma)$ through the expressions
\begin{eqnarray}
{\cal B}^i(\vec x)&=&\left.-i\epsilon^{ijk}\Delta_{jk}(\vec x)W(\gamma)
\right|_{\gamma=0},\label{DefB}\\
{\cal E}^i(\vec x)&=&\left.-i\epsilon^{ijk}\Delta_{jk}(\vec x)
\Omega(\Gamma)\right|_{\Gamma=0},\label{DefE}
\end{eqnarray}
where we have made use of the loop derivative $\Delta_{ij}(x)$ of
Gambini-Trias\cite{GT-em},
\begin{equation}
\delta\sigma^{ij}\Delta_{ij}(\vec x)f(\gamma)\equiv
f(\delta\gamma\circ\gamma)-f(\gamma)\label{LD},
\end{equation}
that measures the change experimented by a loop dependent object $f(\gamma)$
when its argument $\gamma$ is modified by attaching a small plaquette
$\delta\gamma$ of area $\delta\sigma^{ij}$ at the point $\vec x$. In view of
equations \eref{DefB},\eref{DefE}, the Hamiltonian and the other observables
of the theory may be expressed in terms of the basic operators $W$ and
$\Omega$. Equations \eref{AlgDual}, \eref{RealW} and \eref{RealO} are the
basic results of the geometric formulation of massless theories that we are
going to extend to massive cases, with and without topological terms, in the
following sections.

\section{Proca theory in three dimensions}\label{sec3}

In order to preserve gauge invariance, we start from lagrangian of
the Proca model in the St\"ueckelberg form
\begin{equation}
{\cal L}=-\frac 14F_{\mu\nu}F^{\mu\nu}+\frac 12 m^2(A_\mu+\partial_\mu f)
(A^\mu+\partial^\mu f).\label{LagProca}
\end{equation}
It is a trivial matter to see that the equation of motion associated to the
auxiliar field $f$ is nothing but a consistence requisite for the other
equation, which is the relevant one. This reflects the invariance of the
lagrangian density \eref{LagProca} under the gauge transformations
\begin{eqnarray}
\delta A_\mu&=&\partial_\mu\Lambda,\\
\delta f&=&-\Lambda.
\end{eqnarray}
As in the Maxwell case, $f$ may be eliminated by choosing $f=0$. To
incorporate the dual formulation of the theory \eref{LagProca}, we take
the master lagrangian
\begin{equation}
{\cal L'}=m\epsilon^{\mu\nu\lambda}\partial_\mu A_\nu B_\lambda+
\frac{m^2}2\left(B_\mu+\partial_\mu\omega\right)\left(B^\mu+\partial^\mu\omega\right)+
\frac{m^2}2\left(A_\mu+\partial_\mu f\right)\left(A^\mu+\partial^\mu f\right),
\label{proca1}
\end{equation}
which is first order in the Proca field $A_\mu$ and the dual field $B_\mu$.
Besides $f$ we have introduced the St\"ueckelberg field $\omega$, associated
to $B_\mu$, to promote gauge invariance.

It can be seen that \eref{proca1} corresponds to two self-dual models \cite{TPN}
with opposite spins. In fact if we do the change
\begin{eqnarray}
A_\mu&=&\frac1{\sqrt{2}}\left(a_{\mu}^1+a_{\mu}^2\right),\nonumber \\
B_\mu&=&\frac1{\sqrt{2}}\left(a_{\mu}^1-a_{\mu}^2\right),\nonumber \\
f&=&\frac1{\sqrt{2}}\left(f_1+f_2\right),\nonumber \\
\omega&=&\frac1{\sqrt{2}}\left(f_1-f_2\right).
\end{eqnarray}
we will get two decoupled self-dual lagrangians (see equation
\eref{eq:1.2} further) in St\"uckelberg form. Each of them
describe a massive mode with spin +1 for one mode and spin -1 for
the other \cite{DJ}. The invariance under P and T transformations
is accomplished if we exchange the fields $a_{\mu}^1$ and
$a_{\mu}^2$ (and so with the fields $f_1$ and $f_2$). In this
sense we see that the field $B_{\mu}$ behaves as  a pseudovector.

The equations of motion that
results after eliminating the St\"ueckelberg fields are
\begin{eqnarray}
\epsilon^{\mu\nu\lambda}\partial_\nu B_\lambda+mA^\mu&=&0,\label{EdM1}\\
\epsilon^{\mu\nu\lambda}\partial_\nu A_\lambda+mB^\mu&=&0.\label{EdM2}
\end{eqnarray}
By substitution of $B^\lambda$ from \eref{EdM2} into
\eref{proca1} we obtain the Proca lagrangian \eref{LagProca}
(with $f=0$). Doing an analogous procedure with $A^\mu$ from
\eref{EdM1} we obtain the same Proca lagrangian, but this time in
terms of the dual field $B_\mu$. In this sense, one says that the
theory is self dual, and  the master lagrangian ${\cal L'}$ is a
good starting point to explore the geometrical consequences of
this duality.

Let us now quantize the theory. From the ``BF'' term  we directly read the
commutator\cite{Faddeev-Jackiw}
\begin{equation}
\left[A_i(\vec x),\epsilon^{kj}B_j(\vec y)\right]=i\frac 1{m}\delta_i^k\dxy.
\end{equation}
The Hamiltonian results to be
\begin{eqnarray}
H=\int d^2\vec x\left(\frac 1{2m^2}\pi_\omega^2+\frac 1{2m^2}\pi_f2+
\frac {m^2}2(B_i+\partial_i\omega)(B_i+\partial_i\omega)\right.
\nonumber\\
    +\frac{m^2}2(A_i+\partial_i f)(A_i+\partial_i f)
-A_0\left(\pi_f+m\epsilon^{ij}\partial_i B_j\right)\nonumber\\
\left.-B_0\left(\pi_\omega+m\epsilon^{ij}
\partial_i A_j\right)\right),\label{Ham-Pro}
\end{eqnarray}
where $\pi_w$ and $\pi_f$ are the momenta conjugate to $\omega$ and $f$
respectivelly. We did not consider $A_0$ and $B_0$ as canonical variables since
their role as lagrange multipiers is clear. They are associate to the first
class constraints
\begin{eqnarray}
\pi_f+m\epsilon^{ij}\partial_i B_j&\approx&0,\label{v1}\\
\pi_\omega+m\epsilon^{ij}\partial_i A_j&\approx&0,\label{v2}
\end{eqnarray}
that generate the time independent gauge transformations of the
theory. At this point we can compare equation \eref{Ham-Pro} with
the hamiltonian of the Proca theory obtained from the standard
action. Starting from equation \eref{LagProca} with $f=0$, and
following the canonical quantization procedure one obtains the
hamiltonian
\begin{eqnarray}
H_{\hbox{Proca}}&=&\int d^2\vec x\left(\frac 14F_{ij}F_{ij}+
\frac{m^2}2A_iA_i+\frac 12\pi_i\pi_i\right.\nonumber\\
&&\qquad\qquad\qquad\left.+{\frac1{2m^2}}
\left(\partial_i\pi^i\right)^2\right)\label{HProca},
\end{eqnarray}
after solving the second class constraint to eliminate the time component of
the vector field. In equation \eref{HProca}, $A_i$ and $\pi^i$ are
canonically conjugate. On the other hand, in the gauge invariant model
the Hamiltonian $H$ may be expressed as
\begin{eqnarray}
H=\int d^2\vec x\left(\frac 12\left(\epsilon^{ij}\partial_iA_j\right)^2+
\frac 12\left(\epsilon^{ij}\partial_iB_j\right)^2\right.\nonumber\\
\left.+\frac {m^2}2\left(B_i+\partial_i\omega\right)
\left(B_i+\partial_i\omega\right)+
\frac{m^2}2\left(A_i+\partial_i f\right)\left(A_i+\partial_i f\right)\right).
\end{eqnarray}
The equivalence of the two formulations is clear after fixing $f=0$, $\omega=0$
and identifying $\pi^i$ with $m\epsilon^{ij}B_j$.

Examining the first class constraints one realizes that the gauge invariant
combinations that can be formed from the canonical fields $A_i$, $B_i$, $f$
and $\omega$ are $A_i+\partial_if$ and $B_i+\partial_i\omega$. Hence, it is
natural to introduce the non local Wilson like operators
\begin{eqnarray}
W(\gamma_x^{x'})&=&\exp{\left(ie\int_{\gamma_x^{x'}}dx^i
\left(A_i+\partial_i f\right)\right)},\label{3.18}\\
\Omega(\Gamma_y^{y'})&=&\exp{\left(ie\int_{\Gamma_y^{y'}}dx^i
\left(B_i+\partial_i\omega\right)\right),}\label{3.19}
\end{eqnarray}
where $\gamma_x^{x'}$ ($\Gamma_y^{y'}$) is an open curve in $R^2$, starting
at $x$ ($y$) and ending at $x'$ ($y'$) and e is a constant with units
$L^{-\frac 12}$. These operators play the role of the
Wilson Loop and its dual (equations \eref{WL-Maxwell},\eref{WD-Maxwell}) in
the four dimensional Maxwell theory. In virtue of the constraints 
\eref{v1}, \eref{v2} the introduction of non-local operators associated 
with $\pi_f$ and $\pi_{\omega}$ would be redundant. In fact, the exponential 
of $i$ times the integral of $\pi_{\omega}$ over the region of ${\cal R}^2$
bounded by a closed contour $\cal C$ is equivalen to $W(\cal C)$. A similar argument 
holds for $\pi_f$ and $\Omega$.

It is simple to show that the operators
\eref{3.18},\eref{3.19} obey
\begin{equation}
W(\gamma)\Omega(\Gamma)=e^{-i{\frac {e^2}m}N(\gamma,\Gamma)}\Omega(\Gamma)W(\gamma),
\label{DA-Proca}
\end{equation}
where
\begin{eqnarray}
N(\gamma,\Gamma)&=&\int_\gamma dy^i\int_\Gamma dx^j\epsilon^{ij}\dxy,\nonumber \\
&=&\int_{\Gamma} dx^j\epsilon^{ij}T^i(\vec x,\gamma), \nonumber \\
&=&\int_{\gamma} dx^i\epsilon^{ij}T^j(\vec x,\Gamma),
\label{NumCort}
\end{eqnarray}
is the oriented number of intersections between the curves $\gamma$ and
$\Gamma$. This topological quantity obeys the relations
\begin{eqnarray}
N(\gamma,\Gamma)&=&-N(\Gamma,\gamma),\\
N(\gamma_1,\gamma_2\cdot\gamma_3)&=&N(\gamma_1,\gamma_2)+
N(\gamma_1,\gamma_3).
\end{eqnarray}
Ec.\eref{DA-Proca} is the Dual Algebra for the $D=2+1$ Proca
theory. As it was pointed out in the introducction, its
topological "structure constant" involves the intersection index
of the curves that enter in the definition of the non-local
operators, instead of linking numbers, which are the objects that
appear in the DA of masless theories\cite{LL}.

Our next step will be to
realize the DA \eref{DA-Proca} in an appropriate geometrical
representation. To this end we employ the Abelian open path
representation, which has been discussed in
references\cite{LRC,Rolando}. The main features may be summarized
as follows. One groups the piecewise continue (and not
necessarily closed) curves of $R^2$ in equivalence classes
characterized by the equality of their form factors $T(\vec
x,\gamma)$. Then the usual composition of curves turns into a
group product. It is a trivial matter to show that the Abelian
group of loops is a subgroup of the Abelian group of open paths.
Besides the loop derivative eq. \eref{LD}, one can define the
path derivative\cite{LRC,LO}
\begin{equation}
h^i\delta_i(\vec x)\Psi(\gamma)=\Psi(\delta\gamma\circ\gamma)-
\Psi(\gamma),\label{PD}
\end{equation}
which computes the variation of a path-dependent function when an
infinitesimal open path $\delta\gamma_x^{x+h}$ going from $x$ to
$x+h$ ($h\rightarrow 0$) is appended to $\gamma$. It is related to
the loop derivative through the expression
\begin{equation}
\Delta_{ij}(\vec x)=\partial_i\delta_j(\vec x)-\partial_j\delta_i(\vec x).
\end{equation}

The DA \eref{DA-Proca} may be realized onto open-path dependent wave functionals
in the form
\begin{eqnarray}
W(\gamma)\Psi(\gamma_1)&=&\Psi(\gamma\circ\gamma_1),\label{48}\\
\Omega(\Gamma)\Psi(\gamma_1)&=&e^{i{\frac {e^2}m}N(\Gamma,\gamma_1)}
\psi(\gamma_1).\label{49}
\end{eqnarray}
As in the Maxwell case, we have chosen a geometric representation in which
the non local operator associated to the ``direct'' field, i.e., the Wilson
path, produces a ``translation'' in path-space, while that associated to the
dual field is diagonal. One could also interchange these roles. Since the
theory is self-dual, the dual geometric representation results to be a path
representation too.

With the use of the derivative \eref{PD}, the basic local observables of the
theory may be obtained from the non local dual operators
\begin{eqnarray}
A_i+\partial_if& = &\left.-i{\frac {\delta_i(\vec x)}e}W(\gamma)
\right|_{\gamma=0},\\
B_i+\partial_i\omega& = &\left.-i{\frac {\delta_i(\vec
x)}e}\Omega(\Gamma) \right|_{\Gamma=0}.
\end{eqnarray}

As we show in the Appendix, the program developed in this section can also be carried out for 
massive $p$-forms in
  arbitrary dimensions. In $D$ dimensions, Abelian massive theories of $p_1$ and $p_2$-forms are 
dual for $p_1 + p_2 = D-1$.

For instance, the four dimensional Proca theory is dual to the massive Kalb-Ramond model. On the other hand, massive $p$-form theories are associated to generalized Wilson Surfaces $W(\Sigma_p)$, where $\Sigma_p$ is 
an open $p$-surface. Then the Dual Algebra generalizes to
\begin{equation}
W(\Sigma_{p_1})\Omega(\Sigma_{p_2})=e^{-iN(\Sigma_{p_1},\Sigma_{p_2})}\Omega(\Sigma_{p_2})W(\Sigma_{p_1}),
\end{equation}
  where $N(\Sigma_{p_1},\Sigma_{p_2})$ is the intersection index of the open surfaces $\Sigma_{p_1}$ 
and $\Sigma_{p_2}$.

\section{Self-Dual and Topologically Massive theories}\label{sec4}
\subsection{Master Lagrangian and Canonical Quantization}
It is well known that the Topologically Massive \cite{DJT}
\begin{equation}
   \label{eq-1.1}
   S_{TM}=\int d^3x\left(-\frac 14F_{\mu\nu}F^{\mu\nu}+
\frac m{4}\emn F_{\mu\nu}(A)A_\lambda\right),
\end{equation}
and Self-Dual theories \cite{TPN}
\begin{equation}
\label{eq:1.2}
S_{AD}=\int d^3x\left(-\frac m{4}\emn F_{\mu\nu}A_\lambda+
\frac{m^2}2A_\mu A^\mu\right),
\end{equation}
provide locally equivalent descriptions of spin $1$ massive particles in
$2+1$ dimensions \cite{DJ}, although they exhibit different global behaviors
depending on the topological properties of the space time where they are
defined\cite{PR,ST}.

The local equivalence between these models may be viewed by noticing that
they are dual, in the sense that both  may be obtained from the master action
\cite{DJ}
\begin{equation}
   \label{eq:1-1}
   S_M=\frac m{2}\pe{\emn F_{\mu\nu}(A)\left(C_\lambda+
\frac 12A_\lambda\right)+ mC_\mu C^\mu},
\end{equation}
where $F_{\mu\nu}=\partial_{\mu}A_{\nu}-\partial_{\nu}A_{\mu}$. We shall take $g_{\mu\nu}= \hbox{diag}(1, 
-1,-1)$.
The equations of motions, obtained by varying the independent fields
$A_\mu,C_\mu$, are
\begin{eqnarray}
   \label{eq:1-2}
   &&\emn\partial_\nu\left(C_\lambda+A_\lambda\right)=0,\\
   &&\emn \partial_{\nu}A_{\lambda}+ mC^\mu=0\label{eq:1-3}.
\end{eqnarray}
Using equation \eref{eq:1-2} to eliminate $A_\lambda$ in
\eref{eq:1-3} we obtain the equations of motion for the SD model.
In other direction, from equation \eref{eq:1-3} we can eliminate
$C_\mu$ in \eref{eq:1-2} to obtain the equations of of motion of
the TM model. This proves the local classical equivalence.

As in the previous sections, the ``$C^2$'' term spoils gauge invariance.
We remedy this fact by introducing an auxiliary St\"ueckelberg field $\omega$

\begin{eqnarray}
  S_M'=\frac m{2}\int d^3x\left(\emn F_{\mu\nu}(A)\left(C_\lambda+\frac 
12A_\lambda\right)\right.\nonumber\\
+m(C_\mu+\partial_\mu\omega)(C^\mu+\partial^\mu\omega)\Big)\label{eq:1-4}.
\end{eqnarray}
This action is invariant under the simultaneous gauge transformations

\begin{eqnarray}
   \label{eq:1-5}
   \delta A_\mu&=&\partial_\mu\xi,\\
   \delta C_\mu&=&\partial_\mu\zeta,\label{eq:1-6}\\
   \delta\omega&=&-\zeta\label{eq:domega}.
\end{eqnarray}
 From equation (\ref{eq:domega}) we see that the field $\omega$
is  pure gauge, as corresponds to the St\"ueckelberg formulation.

Now we apply the canonical procedure of quantization to the master model.
First, we decompose the action \eref{eq:1-4} into  spatial and temporal parts

\begin{eqnarray}
   S_M'&=&\int d^3x\left(m\varepsilon^{ij}F_{0i}\left(C_j+
\frac 12A_j\right)\nonumber\right.\\
&&+\frac m{2}\varepsilon^{ij}F_{ij}\left(C_0+\frac 12 A_0\right)+
\frac m{2}(C_0+\dot\omega)^2\nonumber\\
&&\left.-\frac m{2}(C_i+\partial_i\omega)
(C_i+\partial_i\omega)\right)\label{eq:1-7}.
\end{eqnarray}
The canonical momenta conjugate to the dinamical variables $A_i$, $C_i$
and $\omega$ are
\begin{eqnarray}
   \label{eq:1-8}
   \pi^i_A&=&m\varepsilon^{ij}\left(C_j+\frac 12 A_j\right),\\
   \pi^i_C&=&0,\label{eq:1-9}\\
   \pi_\omega&=&m(C_0+\dot\omega)\label{eq:1-10}.
\end{eqnarray}
We consider the fields $A_0$ and $C_0$ as non dynamical. They will appear in the
next step as Lagrange multipliers. Equations \eref{eq:1-8} and \eref{eq:1-9}
are just primary constraints among the phase space variables
\begin{eqnarray}
   \label{eq:1-11}
   \psi_i^A&\equiv& \pi^i_A-m\varepsilon^{ij}\left(C_j+\frac 12 A_j\right)\approx 0,\\
   \psi_i^C&\equiv&\pi^i_C\approx 0
\end{eqnarray}
while equation \eref{eq:1-10} allows us to obtain the velocities associated
to $\omega$. Thus, the Hamiltonian on the manifold defined by the primary
constraints is given by
\begin{equation}
   \label{eq:1-13}
   H=\int d^2\vec x\left(\frac1{2m^2}\pi_{\omega}^2+\frac{m^2}{2}(C_i+\partial_i\omega)^2
+A_0\theta_1+C_0\theta_2\right),
\end{equation}
where
\begin{equation}
   \label{eq:1-14}
   \theta_1\equiv-m\varepsilon^{ij}\partial_i(A_j+C_j);
\quad\theta_2\equiv-m\varepsilon^{ij}\partial_iA_j -\pi_\omega.
\end{equation}
Following the scheme of quantization of Dirac, we extend the
Hamiltonian to the whole phase space:
\begin{eqnarray}
     \tilde H&=&\int d^2\vec x\left(\frac1{2m^2}\pi_{\omega}^2+
\frac{m^2}{2}(C_i+\partial_i\omega)(C_i+\partial_i\omega)\right.\nonumber\\
&&\left.+A_0\theta_1+C_0\theta_2+\lambda_A^i\psi_i^A+\lambda_C^i\psi_i^C\right).\label{eq:1-15}
\end{eqnarray}
At this point we observe that the variables $A_0$ and $C_0$ are
the Lagrange multipliers associated to the ``secondary''
constraints $\theta_1$ and $\theta_2$, respectively. Now, we
define the Poisson Brackets among the canonical variables by
\begin{eqnarray}
   \label{eq:1-16}
   \left\{A_i(\vec x),\pi^j_A(\vec y)\right\}&=&\delta_i^j\dxy\\
   \left\{C_i(\vec x),\pi^j_C(\vec y)\right\}&=&\delta_i^j\dxy\\
   \left\{\omega(\vec x)\pi_\omega(\vec y)'\right\}&=&\dxy
\end{eqnarray}
(the remaining Poisson brackets vanish) and proceed to require the preservation
in time of the contraints, taking $\tilde H$ as the generator of time
translations. This leads to determine the Lagrange multipliers associated
to the primary constraints \eref{eq:1-8} and \eref{eq:1-9}
\begin{eqnarray}
   \label{eq:1-17}
   \lambda_A^i=\partial_iA_0+m\varepsilon^{ij}(C_j+\partial_j\omega),\\
   \lambda_C^i=\partial_iC_0-m\varepsilon^{ij}(C_j+\partial_j\omega),
\end{eqnarray}
and it is seen that no further secondary constraints arise. Substituting
the multipliers into the Hamiltonian yields
\begin{eqnarray}
   \label{eq:1-19}
   \tilde H =\int d^2\vec x\left(\frac1{2m^2}{\pi_{\omega}^2}+
\frac{m^2}{2}(C_i+\partial_i\omega)(C_i+\partial_i\omega)+
A_0\vphi_1\right.\nonumber\\
+C_0\vphi_2+\mu\varepsilon^{ij}(C_j+\partial_j\omega)\psi_i^A\nonumber\\
\left.-\mu\varepsilon^{ij}(C_j+\partial_j\omega)\psi_i^C\right),
\end{eqnarray}
where
\begin{eqnarray}
   \label{eq:1-20}
   \vphi_1&\equiv&-\partial_i\psi_i^A+\theta_1=
-\partial_i\pi^i_A-\frac m{2}\varepsilon^{ij}\partial_iA_j,\nonumber\\
\vphi_2&\equiv&-\partial_i\psi_i^C+\theta_2=
-\partial_i\pi^i_C-\pi_{\omega}-m\varepsilon^{ij}\partial_iA_j,
\end{eqnarray}
result to be the first class constraints of the theory.

The matrix associated to the second class constraints
$\Psi_a\equiv(\psi_i^A,\psi_i^C)$ may be written down as
\begin{equation}
   \label{eq:1-21}
   C_{ab}(\vec x,\vec y)\equiv\left\{\Psi_a(\vec x),\Psi_b(\vec y)\right\}=
-m\varepsilon^{ij}\left(
     \begin{array}{ll}
     1&1\\
     1&0
     \end{array}\right)\dxy.
\end{equation}
Its  inverse is given by
\begin{eqnarray}
   C^{-1}_{ab}(\vec x,\vec y)&\equiv&\left\{\Psi_a(\vec x),
\Psi_b(\vec y)\right\}^{-1}\nonumber\\
                             &=&-\frac{1}m\varepsilon^{ij}\left(
     \begin{array}{cc}
     0&1\\
     1&-1
     \end{array}\right)\dxy\label{eq:1-22},
\end{eqnarray}
and allows us to define the Dirac brackets
\begin{equation}
\db FG\equiv\left\{F,G\right\}
-\int d^2\vec xd^2\vec y\left\{F,\Psi_a(\vec x)\right\}
C^{-1}_{ab}(\vec x,\vec y)\left\{\Psi_b(\vec y),G\right\}.\nonumber\\
\end{equation}
Recalling that Dirac brackets are consistent with  second class
constraints, we can eliminate $\pi^i_A$ and  $\pi^i_C$ from now
on, employing the constraints $\Psi_a$. The brackets between the
reduced phase space variables are then

\begin{eqnarray}
\db{A_i(\vec x)}{A_j(\vec y)}&=&0\label{DBA-1}\\
\db{A_i(\vec x)}{C_j(\vec y)}&=&\frac1{m}\epsilon^{ij}\dxy\\
\db{C_i(\vec x)}{C_j(\vec y)}&=&-\frac1{m}\epsilon^{ij}\dxy\\
\db{\omega(\vec x)}{\pi_\omega(\vec y)}&=&\dxy\label{DBA-Ult}.
\end{eqnarray}
Once the phase space has been reduced, the first class constraints become
\begin{equation}
\theta_1\approx 0\qquad,\qquad\theta_2\approx 0.
\end{equation}
and it can be seen that the time independent gauge transformations
generated by these constraints on the reduced phase space variables are
given by
\begin{eqnarray}
\delta A_i(\vec x)&=&\partial_i\xi,\\
\delta C_i(\vec x)&=&\partial_i\zeta,\\
\delta\omega(\vec x)&=&-\zeta,\\
\delta\pi_\omega&=&0,
\end{eqnarray}
as expected.

The next step in the quantization procedure is to promote the
fields to operators acting on a Hilbert space $\cal H$, obeying
commutation relations given by $i\db\cdot\cdot$, and ask the
physical vectors $\ket\psi$ to belong to the kernel of both first
class constraint operators: $\theta_1\ket\psi=0$ and
$\theta_2\ket\psi=0$ . The basic observables (in the sense of
Dirac), from which all relevant gauge invariant information of
the theory can be recovered, are the operators
$-\epsilon^{ij}\partial_i A_j$, $C_i+\partial_i \omega$ and
$\pi_\omega$. It is then natural, within the spirit of the
previous sections, to introduce the non-local operators

\begin{eqnarray}
W(\cal C)&\equiv&\exp{\left(ie\oint_{\cal C}A_idx^i\right)},\label{WL}\\
\Omega(\gamma_x^{x'})&\equiv&\exp{\left(ie\int_{\gamma_x^{x'}}
\left(C_i+\partial_i\omega\right)dx^i\right)}.\label{Wpath}
\end{eqnarray}
In this expression, $e$ is a constant with dimensions
$L^{-\frac12}$. The Wilson Loop  ($W$) and Wilson Path ($\Omega$)
operators, depend on the closed and open paths  $\cal C$ and
$\gamma$ respectivelly. It is easy to see from equations
\eref{DBA-1}-\eref{DBA-Ult} that these operators obey
\begin{eqnarray}
W({\cal C})W({\cal C'})&=&W({\cal C'})W({\cal C})\label{dual-al1}\\
W({\cal C})\Omega(\gamma)&=&e^{-i\frac{e^2}m N({\cal C},\gamma)}
\Omega(\gamma)W({\cal C})\label{dual-al2}\\
\Omega(\gamma)\Omega(\Gamma)&=&e^{i\frac{e^2}m N(\gamma,\Gamma)}
\Omega(\Gamma)\Omega(\gamma)\label{dual-al3},
\end{eqnarray}
where $N(\gamma,\Gamma)$, the oriented number of intersections
between $\gamma$ and $\Gamma$, was defined in equation
\eref{NumCort}.

One could also introduce a non local operator associated to $\pi_\omega$:
the exponential of $i$ times the integral of $\pi_\omega$ over the region 
of ${\cal R}^2$ bounded by a closed contour $\cal C$. However, in virtue of 
the first class constraint $\theta_2\approx 0$, this operator would be just another
representation for the Wilson Loop \eref{WL} when restricted to the physical
space of states, so we do not gain anything with its introduction.

It should be remarked that, as in the previous cases, the local
gauge invariant operators may be obtained from the non-local ones.
In fact, the local operator $C_i+\partial_i\omega$ can be
recovered from the Wilson path by considering an infinitesimal
open path $\delta\gamma$, i.e.
\begin{equation}
\Omega(\delta\gamma)=1+ie\delta\gamma^i\left(C_i+\partial_i\omega\right)+
O(\delta\gamma^2).\label{C-O}
\end{equation}
On the other hand, if $\delta\gamma$ is closed, we have
\begin{equation}
\Omega(\delta\gamma)=1+ie\delta\sigma^{ij}\tilde F_{ij}+O(\delta\sigma^2),
\end{equation}
where $\tilde F_{ij}\equiv\partial_iC_j-\partial_jC_i$. In a similar way,
the local gauge invariant operator $F_{ij}=\partial_iA_j-\partial_jA_i\equiv\epsilon_{ij}B$
and the Wilson Loop are related through
\begin{equation}
W(\delta\gamma)=1+ie\delta\sigma^{ij}\epsilon_{ij}B+O(\delta\sigma^2),\label{B-W}
\end{equation}
where $\delta\gamma$ is an infinitesimal loop. From the latter expansions it is
straight forward to see that
\begin{eqnarray}
C_i+\partial_i\omega&=&\left.-\frac{i}{e}\delta_i(\vec x)\Omega(\gamma)
\right|_{\gamma=0},\\
B&=&\left.-\frac{i}{2e}\epsilon^{ij}\Delta_{ij}(\vec x)W(C)
\right|_{C=0}.
\end{eqnarray}

Finally, the evolution of the physical states is governed by the Schr\"odinger equation
\begin{equation}
i\frac d{dt}\ket{\psi(t)}=\int d^2\vec x\left(\frac12B^2 +
m^2(C_i+\partial_i\omega)^2\right)\ket{\psi(t)}.
\end{equation}
\subsection{Geometrical Representation and the Dual Algebra}

The algebra of non local operators \eref{dual-al1}-
\eref{dual-al3} may be realized on the space of open-path
dependent functionals $\psi(\gamma)$ of section \ref{sec3}, if we
prescribe
\begin{eqnarray}
W(\cal C)\psi(\gamma)&\equiv&\exp{\left(i\frac{e^2}m
N({\cal C,\gamma})\right)}\psi(\gamma)\label{realization1}\\
\Omega(\Gamma)\psi(\gamma)&\equiv&
\exp{\left(-i\frac{e^2}{2m}N(\Gamma,\gamma)\right)}
\psi(\Gamma\circ\gamma).\label{realization2}
\end{eqnarray}
For instance, using eq. \eref{realization2} one has
\begin{eqnarray}
\Omega(\Gamma)\Omega(\Gamma')\psi(\gamma)&=&\Omega(\Gamma)
\left[e^{-i\frac{e^2}{2m}N(\Gamma',\gamma)}\psi(\Gamma'\circ\gamma))
\right]\nonumber\\
                                          &=&e^{-i\frac{e^2}{2m}
N(\Gamma,\gamma)}e^{-i\frac{e^2}{2m}
N(\Gamma',\Gamma\circ\gamma)}\psi(\Gamma\circ\Gamma'\circ\gamma)\nonumber\\
                                          &=&e^{i\frac{e^2}m
N(\Gamma,\Gamma')}\Omega(\Gamma')\Omega(\Gamma)\psi(\gamma),
\end{eqnarray}
in agreement with eq. \eref{dual-al3}. Besides realizing the
non-local algebra, we have to consider the restrictions that the
first class constraints impose onto the path dependent states. The
constraint $\theta_2$ is automatically satisfied if the non local
operator associated to $\pi_\omega$ (see comment after
\eref{dual-al3}) is realized as (essentially) the Wilson Loop $W$.
So it remains to study the constraint $\theta_1$ :
\begin{equation}
\epsilon^{ij}\partial_i(A_j+C_j)\approx 0,
\end{equation}
which may be imposed onto the states as
\begin{eqnarray}
\exp{\left(ie\oint_{{\cal C}}(\hat A_i+\hat C_i)dx^i\right)}\ket\psi
&=&\exp\left(-\frac12\oint_{\cal C}\oint_{\cal C}dx^idy^j\con{\hat A_i(\vec x)}
{\hat C_j(\vec y)}\right)W({\cal C})\Omega({\cal C})\ket\psi,\nonumber \\
&=&W({\cal C})\Omega({\cal C})\ket\psi,\nonumber \\
&=&\ket\psi, \label{Vin-NL}
\end{eqnarray}
or, in other words,
\begin{equation}
W({\cal C}) \approx \Omega(\overline{\cal C}),  \label{phys}
\end{equation}
within the physical sector of the Hilbert Space. To obtain this
result we used that $N({\cal C}_1 , {\cal C}_2)= 0$ for closed
paths ${\cal C}_1 , {\cal C}_2$. Equation \eref{phys} states that
instead of a pair of "Wilson" operators it suffices with just one
of them, namely, $\Omega(\gamma)$ , that simultaneously plays the
role of "coordinate" and "momentum". That $N({\cal C}_1 , {\cal
C}_2)$ vanishes when the curves are closed also matters to see
that when eq. \eref{phys} holds (i.e., on the physical sector),
equation \eref{dual-al3} already implies eqs. \eref{dual-al1} and
\eref{dual-al2}. Therefore, it is just eq. \eref{dual-al3} which
corresponds to the Dual Algebrae of the previous sections
(equations \eref{AlgDual} and  \eref{DA-Proca}), with which it
should be compared.

So far, it remains to study how eq. \eref{phys} restricts the
space of states. Combining equation \eref{phys} with
\eref{realization1} and \eref{realization2} one obtains
\begin{equation}
\psi({\cal C}\circ\gamma)=e^{-i\frac{e^2}{2m}N({\cal C},\gamma)}
\psi(\gamma)\label{vin-int}.
\end{equation}
Taking ${\cal C}$ to be an infinitesimal closed path we can see
that this equation is just the non-local version of the
differential constraint recently obtained in a study of the
path-space quantization of the Maxwell-Chern-Simons theory
\cite{LO}
\begin{equation}
\left(\rho(\vec x,\gamma)-i\frac m
{e^2}\epsilon^{ij}\Delta_{ij}(\vec x) \right)\psi(\gamma)=0,
\label{ppp}
\end{equation}
where
\begin{eqnarray}
  \rho(\vec x,\gamma) & \equiv & -\partial_i T^i(\vec
  x,\gamma)\nonumber\\
  &=& -\sum_s (\delta^2(\vec x - \vec {\beta}_s) -\delta^2(\vec x - \vec
  {\alpha}_s)),\label{ro}
\end{eqnarray}
is a functional that depends on the boundary
$(\vec\alpha,\vec\beta\,)$ of the path $\gamma$. Since the path
may comprises several pieces $s$, both $\vec\alpha$ and
$\vec\beta$ denote the set of starting and ending points,
respectivelly. It can be said then that  $\rho(\vec x,\gamma)$ is
the "density of extremes" of the path $\gamma$. The solution to
\eref{ppp} was found to be\cite{LO}
\begin{equation}
\psi(\gamma)=e^{i\chi(\gamma)}\Phi(\vec\alpha,\vec\beta),\label{oswaldo}
\end{equation}
where $\Phi(\vec\alpha,\vec\beta\,)$ is an arbitrary functional of
the boundary $(\vec\alpha,\vec\beta\,)$ and
\begin{eqnarray}
\chi(\gamma)&=&-i\frac{e^2}{4\pi m}\sum_s\int_\gamma
dx^i\epsilon^{ij}\left(\frac{(x^j-\beta^j_s)}{|\vec
x-\vec\beta_s|^2}
-\frac{(x^j-\alpha^j_s)}{|\vec x-\vec\alpha_s|^2}\right)\nonumber\\
&=&-i\frac{e^2}{4\pi m}\Delta\Theta(\gamma),\label{chi}
\end{eqnarray}
with $\Delta\Theta(\gamma)$ being the algebraic sum of the angles
subtended by the pieces of $\gamma$, measured from their ending
points $\vec\beta$, minus that measured from their starting points
$\vec\alpha$.

  The path-dependent
function $\chi(\gamma)$ is ill defined due to the ambiguous
definition of the angle subtended by a path when it is measured
from their own ending points. In fact, when the point from which
the angle is measured coincides with one of the extremes of the
path, one loses the straight line connecting that point with the
extreme, which would serve as a reference to compute the desired
angle. We can replace that fidutial straight line by the tangent
to the path at the problematic point. For instance, if we want to
compute the angle subtended by the path $\gamma$, given as a map
from the interval $[0,1]$ to $R3$, measured from its starting
point $\vec y(0)=\vec\alpha$, we can take the prescription
\begin{equation}
\Theta(\gamma,\vec\alpha)\equiv\lim_{a\rightarrow0^+}\int_a^1dt\frac{dy^i(t)}{dt}
\epsilon^{ij}\frac{(y^j(t)-\alpha^j)}{|\vec y-\vec\alpha|^2}.
\end{equation}
It may be seen that this prescription is consistent with the fact
that $\chi(\gamma)$ must be a path-dependent function, and not
merely a curve-dependent one.

It is worth noticing, from eq. \eref{oswaldo}, that the path
dependence of the wave functionals is realized through the
boundary points of the paths, and through the way they wind
around these points. Hence, we see that in this case not only the
DA shows a topological character, but also the geometrical
representation of the algebra carries a topological content.

Equations \eref{C-O}-\eref{B-W} together with the realization
\eref{realization1} and \eref{realization2} for the DA allow us to
see that the gauge invariant operators $B(\vec x)$ and $C_i(\vec
x)+\partial_i\omega(\vec x)$ are realized as

\begin{eqnarray}
B(\vec x)&\rightarrow&-i\frac{e^2}m\rho(\vec x,\gamma)\\
C_i(\vec x)+\partial_i\omega(\vec x)&\rightarrow&-i{\cal
D}_i(\vec x) \equiv -i\delta_i(\vec
x)-\frac{e^2}{2m}\epsilon^{ij}T^j(\vec x,\gamma),
\end{eqnarray}
whose action on gauge invariant functionals can be seen to
respect the form  given in eq.\eref{oswaldo} \cite{LO}. The same
is then true for the non-local operator $\Omega({\gamma})$, in
view of its definition \eref{Wpath}.

Now, let us quote the path-representation expressions for
the Poincare generators of the theory:

\begin{eqnarray}
H&=&\frac{m^2}{e^2}\int d^2\vec x\left[-\Delta_{ij}(\vec x)\Delta_{ij}(\vec x)
-\frac 12{\cal D}_i(\vec x){\cal D}_i(\vec x)\right]\label{Real-H}\\
P^i&=&\frac{m}{2e^2}\int d^2\vec x\epsilon^{jk}\Delta_{jk}(\vec x)i{\cal D}_i(\vec x),\\
J&=&\frac{m}{e^2}\int d^2\vec
x\epsilon^{ij}x^i\epsilon^{kl}\Delta_{kl}(\vec x){\cal D}_j(\vec
x)
\end{eqnarray}
It can be shown that the operator $P^i$ ($J$) generates rigid
translations (rotations) of the path $\gamma$ appearing in the
argument of the wave functional $\psi(\gamma)$, as it should be
expected. Since $\chi(\gamma)$ is invariant under both
translations and rotations, the above results does not contradict
the fact that $P^i$ and $J$ are gauge invariant operators. In
other words, one has, for an infinitesimal translation along $u^i$
\begin{equation}
\left(1+u^iP^i\right)\psi(\gamma)=e^{i\chi(\gamma)}
\left(1+u^iP^i\right)\Phi(\vec\alpha,\vec\beta).
\end{equation}
Thus, $P^i$ translates the boundary $(\vec\alpha,\vec\beta)$ of
the path while maintaining the form of the wave functional given
by \eref{oswaldo}, which is dictated by gauge invariance. A
similar argument holds for infinitesimal rotations.

\section{Discussion}
We have studied the duality symmetry between massive Abelian
$p-$forms, with and without topological terms, from the point of
view of the geometrical representations that, in each case,
generalize the loop representation of Maxwell theory. We found
that in the cases without topological terms, and within the
physical sector of the Hilbert space, the canonical algebra of
local operators can be translated into a non local albegra of a
pair of gauge invariant operators, that exhibit an interesting
geometrical content, and that is characterized by a topological
quantity, namely, the intersection index between the geometrical
objects that constitute the argument of the gauge invariant
operators. This algebra, the Dual Algebra, may be realized in a
basis of wave functionals depending on open paths, or
$p-$surfaces, according to the rank of the forms involved. In
general, for any pair of dual theories, there is also a pair of
dual geometric representations. This situation degenerates in the
case of self-duality, since then the ``direct'' and the ``dual''
theories are equivalent.

Regarding the study of the TM and SD case theories in $2+1$
dimensions, we found that, as in the Proca model in $2+1$
dimensions, the topological quantity that characterizes the DA is
the number of intersections of two paths. However, unlike the
Proca's case, these paths are different arguments of the same
operator. Another important difference is that in the TM and SD
models, the open paths involved fall into equivalence classes
labeled by their boundary $\partial\gamma$ and their winding
properties described by $\Delta\Theta(\gamma)$. One could say
that in this case the geometric representation in one of ``rubber
bands with fixed ends'', rather than a path representation.

The results of this study could contribute to put both massless
and massive Abelian gauge theories under a common scope, regarding
their geometrical properties. It remains to explore whether or
not these ideas find a suitable extension to the non-Abelian case. Also, it 
would be interesting to study how the equivalence between the Proca model and 
two selfdual models with opposite spins is manifested in the geometrical representation.

\appendix
\section{Conventional Massive Theories}\label{Ap-A}
In this appendix we discuss briefly how to extend the results of
section \ref{sec3} to the general case of duality between massive
Abelian forms $A_{\mu_1\cdots\mu_p}$ and
$B_{\mu_1\cdots\mu_{D-p-1}}$, for $p=0,1,\cdots,D-1$, in $D$
spacetime dimensions. We start from the St\"uckelberg form of the
master action, i.e.
\begin{eqnarray}
I_M\sup{p,D}&=&\int d^Dx\left(\frac{g(-1)^{pD}}{(p+1)!(D-p-1)!}\epsilon^{\mu_1\cdots\mu_{p+1}\nu_1
\cdots\nu_{D-p-1}}F_{\mu_1\cdots\mu_{p+1}}(A)B_{\nu_1\cdots\nu_{D-p-1}} -\right.\nonumber\\
&&\qquad\qquad-\frac{g(-1)^p}{2(D-p-1)!}(B_{\mu_1\cdots\mu_{D-p-1}}+
F_{\mu_1\cdots\mu_{D-p-1}}(C))^2 -\nonumber\\
&&\qquad\qquad\qquad\left.-\frac{g(-1)^p\mu^2}{2p!}(A_{\mu_1\cdots\mu_p}+
F_{\mu_1\cdots\mu_p}(\omega))^2\right),\label{general}
\end{eqnarray}
with $F(f) \equiv df$ for any form $f$. Here, $\omega$ and $C$
are auxiliary  St\"uckelberg $p-1$ and $D-p-2$ forms
respectivelly. From the  space-time decomposition of the master
action \eref{general}, which is  given by
\begin{eqnarray}
I_M\sup{p,D}&=&\int d^Dx\left({\frac 1{2(D-p-2)!}}
(B_{0i_1\cdots i_{D-p-2}}+
F_{0i_1\cdots i_{D-p-2}}(C))^2\right.\nonumber\\
             &&-{\frac 1{2(D-p-1)!}}(B_{i_1\cdots i_{D-p-1}}+
F_{i_1\cdots i_{D-p-1}}(C))^2\nonumber\\
             &&+{\frac{\mu^2}{2(p-1)!}}
(A_{0i_1\cdots i_{p-1}}+
F_{0i_1\cdots i_{p-1}}(\omega))^2-{\frac{\mu^2}{2p!}}
(A_{i_1\cdots i_p}+F_{i_1\cdots i_p}(\omega))^2\nonumber\\
             &&+{\frac{g(-1)^{pD}}{p!(D-p-1)!}}
\epsilon^{i_1\cdots i_pj_1\cdots j_{D-p-1}}
\dot A_{i_1\cdots i_p}B_{j_1\cdots j_{D-p-1}}\nonumber\\
             &&-{\frac{g(-1)^{pD}}{(p-1)!(D-p-1)!}}
\epsilon^{i_1\cdots i_pj_1\cdots j_{D-p-1}}
\partial_{i_2} A_{0i_2\cdots i_{p-1}}B_{j_1\cdots j_{D-p-1}}\nonumber\\
             &&\left.-{\frac{g(-1)^{pD}}{p!(D-p-2)!}}(-1)^p
\epsilon^{i_1\cdots i_{p+1}j_1\cdots j_{D-p-2}}\partial_{i_1}
A_{i_2\cdots i_{p+1}}B_{0j_1\cdots j_{D-p-2}}\right),\nonumber\\
\end{eqnarray}
we can read the fundamental Poisson bracket
\begin{equation}
\left\{A_{i_1\cdots i_p}(\vec x),B_{j_1\cdots j_{D-p-1}}
(\vec y)\right\}=
g\epsilon^{i_1\cdots i_pj_1\cdots j_{D-p-1}}
\delta\sup d(\vec x-\vec y),\label{Alg-MasLoc}
\end{equation}
and obtain the Hamiltonian as
\begin{eqnarray}
H_M\sup{p,D}&=&\int d^d\vec x\left\{\frac 1{2(D-p-2)!}\left(\pi_C^{i_1\cdots i_{D-p-2}}\right)^2+
\frac 1{2(p-1)!}
\left(\pi_\omega^{i_1\cdots i_{p-1}}\right)^2\right.\nonumber\\
            &&+\frac 1{2(D-p-1)!}
\left(B_{i_1\cdots i_{D-p-1}}+F_{i_1\cdots i_{D-p-1}}(C)\right)^2+
\frac 1{2p!}\left(A_{i_1\cdots i_p}+
F_{i_1\cdots i_p}(\omega)\right)^2\nonumber\\
            &&+B_{0i_1\cdots i_{D-p-2}}
\Theta_1^{i_1\cdots i_{D-p-2}}+
C_{0i_1\cdots i_{D-p-3}}
\Theta_2^{i_1\cdots i_{D-p-3}}+
\omega_{0i_1\cdots i_{p-2}}\Theta_3^{i_1\cdots i_{p-1}}\nonumber\\
            &&\left.+A_{0i_1\cdots i_{p-1}}
\Theta^{i_1\cdots i_{p-1}}\right\}.
\end{eqnarray}
In this equation, the quantities
\begin{eqnarray}
\Theta_1^{i_1\cdots i_{D-p-2}}&=&-\frac 1{(D-p-2)!}
\left[\pi_C^{i_1\cdots i_{D-p-2}}+\frac{g(-1)^{p(D+1)}}{p!}
\epsilon^{j_1\cdots j_{p+1}i_1\cdots i_{D-p-2}}
\partial_{j_1}A_{j_2\cdots j_{p+1}}\right]\nonumber\\
\Theta_2^{i_1\cdots i_{D-p-3}}&=&-\frac 1{(D-p-3)!}\partial_i\pi_C^{ii_1\cdots i_{D-p-3}},\nonumber\\
\Theta_3^{i_1\cdots i_{p-2}}&=&
-\frac 1{(p-2)!}\partial_i\pi_\omega^{ii_1\cdots i_{p-2}},\\
\Theta_4^{i_1\cdots i_{p-1}}&=&
\frac 1{(p-1)!}\left[\pi_\omega^{i_1\cdots i_{p-1}}+
\frac{g(-1)^{p(D+1)}}{(D-p-1)!}
\epsilon^{i_1\cdots i_{p-1}j_1\cdots j_{D-p}}
\partial_{j_1}B_{j_2\cdots j_{D-p}}\right],
\end{eqnarray}
are the secondary first class constraints associated to the
Lagrange multipliers $B_0,C_0,\omega_0$ y $A_0$, respectively.

The non-local and gauge-invariant "Wilson operators" of this
theory are
\begin{eqnarray}
W(\Sigma_p)&=&\exp{\left(i\int_{\Sigma_p}(A+d\omega)\right)},\\
\Omega(\Sigma_{D-p-1})&=&\exp{\left(i\int_{\Sigma_{D-p-1}}\!\!\!\!\!\!\!\!(B+dC)\right)}.
\end{eqnarray}
They obey the Dual Algebra
\begin{equation}
W(\Sigma_p)\Omega(\Sigma_{D-p-1})=\exp{\left(-iN(\Sigma_p,\Sigma_{D-p-1})\right)}\Omega(\Sigma_{D-p-1})W(\Sigma_p),
\end{equation}
where $N(\Sigma_p,\Sigma_{D-p-1})$ is the oriented number of
intersection between the hypersurfaces $\Sigma_p$ and
$\Sigma_{D-p-1}$.
This model admits two dual geometric
representations. In one of them, $W(\Sigma_p)$ appends a
$p-$surface $\Sigma_p$ to the argument of the surface-dependent
functional $\Psi({\Sigma_p '})$ on which it acts, while
$\Omega(\Sigma_{D-p-1})$ counts ($i$ times the exponential of)
how many times $\Sigma_p '$ and $\Sigma_{D-p-1}$ intersect each
other. In the "dual" geometric representation, on the other hand,
these roles are interchanged: $\Omega$ appends $ \Sigma_{D-p-1}$
surfaces while $W$ counts intersection numbers. This result
should be compared with the $2+1$ case discussed in section
\ref{sec3}.

\noindent {\bf Acknowlegment}\\
This work is supported by Project G-2001000712 of FONACIT.


\begin{thebibliography}{25}
\bibitem{Townsend:1981nu}
P.~K.~Townsend,
in {\it C81-02-18.8}
CERN-TH-3067
{\it Lecture given at 18th Winter School of Theoretical Physics, Karpacz, Poland, Feb 18 - Mar 3, 1981}.

\bibitem{DFNS} S.~E.~Hjelmeland and U.~Lindstrom,
hep-th/9705122.

\bibitem{LL} L.~Leal,
Mod.\ Phys.\ Lett.\ A {\bf 11}, 1107 (1996)
[hep-th/9603006].

\bibitem{Kramers:1941kn}
H.~A.~Kramers and G.~H.~Wannier,
Phys.\ Rev.\  {\bf 60}, 252 (1941).

\bibitem{tHooft} G.~'t Hooft,
Nucl.\ Phys.\ B {\bf 153}, 141 (1979).

\bibitem{AlgNL-Maxwell}
M.~B.~Halpern,
Phys.\ Rev.\ D {\bf 19}, 517 (1979).

\bibitem{Marino:1992rq}
E.~C.~Marino and J.~E.~Stephany,
Int.\ J.\ Mod.\ Phys.\ A {\bf 7}, 171 (1992).

\bibitem{TPN} P.~K.~Townsend, K.~Pilch and P.~van Nieuwenhuizen,
Phys.\ Lett.\  {\bf 136B}, 38 (1984)
[ {\bf 137B}, 443 (1984)].

\bibitem{DJT} S.~Deser, R.~Jackiw and S.~Templeton,
Ann. Phys.\  {\bf 140}, 372 (1982).

\bibitem{DJ} S.~Deser and R.~Jackiw,
Phys.\ Lett.\ B {\bf 139}, 371 (1984).

\bibitem{St}
E.~C.~St\"ueckelberg,
Helv.\ Phys.\ Acta {\bf 15}, 23 (1942).

\bibitem{Dirac}P.A.M.~Dirac, {\it Lectures on Quantum Mechanics}, Belfer Graduate
School, Yeshiva University, New York,1964.

\bibitem{Faddeev-Jackiw} L.~Faddeev and R.~Jackiw,
Phys.\ Rev.\ Lett.\  {\bf 60}, 1692 (1988).

\bibitem{GT-em} R.~Gambini and A.~Trias,
Phys.\ Rev.\ D {\bf 22}, 1380 (1980);
D {\bf 23}, 553 (1981);
D {\bf 27}, 2935 (1983);
X.~Fustero, R.~Gambini and A.~Trias,
Phys.\ Rev.\ D {\bf 31}, 3144 (1985).

\bibitem{NC:Max} C.~di Bartolo, F.Nori, R.~Gambini and A.~Trias,
Lett.\ Nuovo Cim.\  {\bf 38}, 497 (1983).

\bibitem{LRC} J.~Camacaro, R.~Gaitan and L.~Leal,
Mod.\ Phys.\ Lett.\ A {\bf 12}, 3081 (1997)
[hep-th/9606121].

\bibitem{Rolando} R.~Gaitan and L.~Leal,
Int.\ J.\ Mod.\ Phys.\ A {\bf 11}, 1413 (1996).


\bibitem{LO} L.~Leal and O.~Zapata,
Phys.\ Rev.\ D {\bf 63}, 065010 (2001)
[hep-th/0008049].


\bibitem{PR} P.~J.~Arias and A.~Restuccia,
Phys.\ Lett.\ B {\bf 347}, 241 (1995)
[hep-th/9410134].

\bibitem{ST} J.~Stephany,
Phys.\ Lett.\ B {\bf 390}, 128 (1997)
[hep-th/9605074].































\end{thebibliography}
\end{document}